\documentstyle[prl,aps]{revtex} 
\tighten
\begin{document}
    
%\draft 
\title{High energy solar neutrinos and p-wave contributions to
$^3$He(p,$\nu$e$^+$)$^4$He}

\author{C.  J.  Horowitz\footnote{email:  charlie@iucf.indiana.edu}} 
\address{Nuclear Theory Center and Dept. of Physics \\ 
Indiana University \\ 
Bloomington, Indiana 47405}

\date{\today} 
\maketitle 
\begin{abstract} High energy solar neutrinos can come from the
hep reaction $^3$He(p,$\nu$e$^+$)$^4$He with a large end point energy of 18.8 MeV.
Understanding the hep reaction may be important for interpreting solar neutrino spectra.
We calculate the contribution of the axial charge transition $^3P_0\rightarrow\,
{}^1S_0$ to the hep thermonuclear S factor using a one-body reaction model involving a
nucleon moving in optical potentials.  Our result is comparable to or larger than
previous calculations of the s-wave Gamow Teller contribution.  This indicates that the
hep reaction may have p-wave strength leading to an enhancement of the 
$S$ factor.
\end{abstract} 

After many years of work on solar neutrinos, experimenters are now searching for proof
of new neutrino physics that is independent of solar models.  Super-Kamiokande is
looking for distortions in the shape of the $^8$B spectrum from neutrino
oscillations\cite{superK}.  Indeed, the ratio of the measured super-K spectrum to the 
expected ${}^8$B spectrum rises at high energies.  Among other possibilities,  this 
could be due to oscillations or to neutrinos from the hep reaction, 
$$p+ {}^3{\rm He} \rightarrow {}^4{\rm He} + \nu_e + e^+.\eqno(1)$$ 
Although rare, hep neutrinos have a higher end point (18.8 MeV) than those from 
$^8$B.  Thus, the interpretation of solar neutrino spectra may depend 
on our understanding of the hep reaction\cite{bahcall,escribano}.

The present estimate for the thermonuclear $S$ factor for Eq.  (1) is small\cite{BP98}
$$S_0=2.3\times 10^{-20}\ {\rm keV-b}\eqno(2)$$ based on the calculations of Carlson et
al.\cite{carlson,schiavilla}.  One would need an $S$ factor some 20 times larger to
explain the super-K data\cite{bahcall}.  However, Carlson et al.  only consider the
contribution of a single partial wave and only keep the axial-three-vector part of the
weak current.  Furthermore, they neglect the radial dependence of the lepton wave
functions.  This dependence could be significant because of the large Q value.

Carlson et al.  find a small $S$ factor because of sensitive cancellations in the wave
function and destructive interference between one-body and meson exchange current
contributions.  Given the very small Gamow-Teller strength it is important to study
other forbidden transitions which may also contribute.

Some earlier work on the hep $S$ factor assumed a relation between radiative capture
${}^3$He($n,\gamma){}^4$He and Eq. (1)\cite{werntz1,werntz2,tegner,wolfs,wervelman}.
However, the weak and electromagnetic currents are very different so this relation may
be unreliable\cite{carlson}.  Werntz and Brennan\cite{werntz2} estimate the 
contributions of p-wave resonances to Eq. (1).  However, we are not aware of any 
nonresonant p-wave calculations.

In this paper we study the axial-charge transition $^3P_0\rightarrow\, {}^1S_0$ which
involves a p-wave initial state.  Our goal is to show, in as simple a way as possible,
that p-waves can compete with the small s-wave strength.  Therefore we focus on a single
partial wave and operator.  To calculate the total $S$ factor one must add coherently the
contribution of several other p-wave transitions, other operators and the original 
s-wave strength to our axial charge result.  We find the axial charge transition 
is comparable to the original s-wave.  Furthermore, our result is based on the 
one-body axial charge.  It is known that meson exchange currents, rather than 
interfering destructively, significantly enhance many axial charge transitions.  
For example, the cross section for near threshold pion production, 
$pp\rightarrow pp\pi^0$, is enhanced by a factor of five 
by meson exchange currents\cite{pions}.

We use a simple one-body model to estimate p-waves.  This model has a nucleon moving in
optical potentials chosen to reproduce bound state properties and p-$^3$He phase shifts.
The model is not expected to be good for the s-wave transition since this involves
sensitive cancellations and small components of the wave function.\footnote{If one
assumes the same optical potential for bound and scattering states, our one-body model
has zero s-wave strength because the bound and scattering states are orthogonal.}
However, our model should provide a first estimate for the p-waves since they involve
large components of the wave functions and do not (appear to) have sensitive
cancellations.  We describe our model, give results for the $^3P_0\rightarrow\, {}^1S_0$
contribution to the $S$ factor and then discuss future experimental and theoretical 
work. Our conclusion is that p-wave contributions could increase the hep $S$ factor.

The $^4$He final state $|\Psi_f>$ is modeled as a neutron moving in a real Wood Saxon
optical potential.  $$|\Psi_f> = {u(r)\over r} Y_{00}(\hat r)\chi_0^0\eqno(3)$$ Here $r$
is the relative coordinate between the nucleon and $^3$He while $\chi_0^0$ is a spin
singlet wave function for the nucleon and $^3$He.  The bound state radial wave 
function $u(r)$ is a solution to the Schrodinger equation for a reduced mass 
$\mu={3\over 4}m$, with $m$ 
the nucleon mass, moving in a potential $V(r)$, $$V(r)=V_0/\bigl[ 1 + {\rm
exp}[(r-R_0)/a]\bigr].\eqno(4)$$ We arbitrarily set $a=0.6$ fm while $R_0=2$ fm was
adjusted to reproduce the charge radius of $^4$He after correcting for the center
of mass and folding in the finite size of the proton.  The strength $V_0=-63.9$ MeV
reproduces the 20.58 MeV neutron separation energy of $^4$He.

The $^3P_0$ initial state $|\Psi_i>$ is modeled as
$$|\Psi_i> = i\sqrt{4\pi}{u^+_1(r)\over r} |{}^3P_0>,\eqno(5)$$ 
with $|{}^3P_0>$ a spin-angle function
and $u^+_1$ a ${}^3P_0$ outgoing scattering wave which we approximate as a p-wave
solution in a coulomb potential of a uniform sphere of radius $R_0=2$ fm plus $V(r)$
given by Eq.  (4).  We use the same $R_0=2$ fm and $a=0.6$ fm as the bound state but
adjust $V_0=-30$ MeV in order to approximately reproduce $p+{}^3$He phase shifts (see
Fig. (1)).

The ${}^3P_0$ contribution to the hep cross section from the axial charge is
\cite{carlson}, $$\sigma = {1\over
(2\pi)^3} {G^2m_e^5f\over v} | <\Psi_f|A_0|\Psi_i> |^2,\eqno(6)$$ with $G=1.151\times
10^{-11}$ MeV$^{-2}$ the Fermi constant, $m_e$ the electron mass, $v$ the p-${}^3$He
relative velocity and $f=2.544\times 10^6$ the lepton phase space.    
Note, the three-vector part of the axial current will also contribute given the 
spatial dependence of the lepton wave functions.  For simplicity we focus on a
single operator.  The one-body axial charge operator $A_0$ is assumed to be, 
$$A_0=-{g_a\sigma\cdot (\vec p + \vec p\, ')\over 2m}\tau_-,\eqno(7)$$ 
with $g_a=1.262$, $p$ the initial and $p'$ the final nucleon momenta
and $\tau_-={1\over 2}(\tau_x-i\tau_y)$ converts a proton into a neutron.

It is a simple matter to evaluate the matrix element using Eqs.  (3, 5 and 7),
$$|<\Psi_f|A_0|\Psi_i>|^2={4\pi g_a^2  \over \mu^2} |\int_0^\infty dr\, 
u(r)[{d\over dr} + {1\over r}] u_1^+(r)|^2.\eqno(8)$$ 
Numerically evaluating Eqs.  (6,8) at 7.5 keV in the
center of mass yields a cross section of $\sigma=5.28\times 10^{-30}$b.  Converting this
to the usual $S$ factor $S=E\sigma {\rm e}^{2\pi\eta}$ with $\eta=2\alpha/v$ yields,
$$S_{{}^3P_0}=1.67\times 10^{-20} {\rm \ keV-b}.\eqno(9)$$ 
Note, because of large Coulomb effects there is little energy dependence to Eq. (9),
$S(E)\propto (1+\eta^{-2})$.
Again Eq. (9) only includes the contribution of the axial charge and a single partial 
wave.  To obtain the total $S$ factor one must add contributions of other p-waves and 
operators and the s-wave strength.  Furthermore, Eq. (9) may be
enhanced by meson exchange currents.

Nevertheless, Eq.  (9) is {\it larger} than the $1.3\times 10^{-20}$ keV-b $S$ factor
originally claimed by Carlson et al.\cite{carlson} and 73\% of the present value,
Eq (2).  We conclude that {\it p-wave contributions may be comparable to the 
Gamow-Teller strength.}  This is a major result of this paper and will be 
discussed below.

How can this axial-charge transition compete with the Gamow-Teller?  First the 
centrifugal barrier's effects are significantly reduced by the strong Coulomb 
interaction.  The ratio of l=1 to l=0 Coulomb wave functions is much larger 
than that for plane waves.  Second, the axial-charge operator is of order 
$v_N/c\approx 0.25$ and the nucleon's velocity $v_N$ is relatively large 
in ${}^4$He because of the large separation energy.   The product of these 
two factors, centrifugal barrier times $v_N/c$, is not very small and can 
compete with a strongly reduced s-wave matrix element.

We now discuss some of the details of the calculation.  We use a very simple one-body
wave function with an optical potential fit to phase shifts.  To explore the sensitivity
to the p-wave optical potential, we consider other values for the strength $V_0$, see
Eq.  (4).  A very conservative choice is to set $V_0=0$ and thus the incoming p-wave
sees only the Coulomb potential.  This is unrealistic because we expect some attractive
nuclear interaction.  Nevertheless, setting $V_0=0$ only reduces the axial-charge
matrix element by 30\%.   We conclude that changes in the p-wave optical potential are
unlikely to significantly reduce the $S$ factor in Eq.  (9).

Alternatively, resonances could significantly enhance the cross section.  If we use the
same value of $V_0=-63.9$ MeV as was used for the bound state then there will be a
strong p-wave resonance and the $S$ factor rises by a factor of 37 to $62.5\times
10^{-20}$ keV-b.  Such a strong resonance is not seen in the p-wave phase shifts.
Therefore this very large $S$ factor is probably unrealistic.  However there could be
contributions from smaller resonances.

We have not explicitly antisymmeterized the incident proton with those bound in $^3$He.
This omission could be important for s-waves.  However we don't expect it to be a large
correction for p-waves.

It is important to repeat our calculation with more realistic microscopic four-body wave
functions.  However, the p-wave transition does not appear to involve sensitive
cancellations.  Furthermore, the axial charge operator can connect the large components
of the wave functions.  Thus the matrix element should not depend strongly on small
components in the wave functons of $^3$He and $^4$He.  Therefore, we expect Eq.  (9) to 
provide a useful first estimate.

Meson exchange currents (MEC) can be important for axial-charge transitions because both
the one-body and MEC are of the same order $v_N/c$.
Pion exchange currents enhance the axial-charge in a number of first 
forbidden beta decays.
In addition, shorter range MEC could also enhance the axial-charge.  In relativistic
models, sigma meson exchange increases the axial charge from order $p/m$ to $p/M^*$
where the nucleon's effective mass $M^*<m$.  This and omega exchange enhance the near
threshold pion production cross section by a factor of five\cite{pions}.  Thus we expect
a significant MEC contribution and we expect it to increase the $S$ factor.

For simplicity we have focused on a single partial wave ${}^3P_0$.  There are a number
of other p-waves which can also contribute such as ${}^1P_1$, ${}^3P_1$ and ${}^3P_2$.
As a very crude estimate we expect these partial waves to each be of the same order of
magnitude as the ${}^3P_0$.  Therefore, it is possible that the total $S$ factor,
involving the coherent sum of several contributions, could be significantly larger 
than Eq. (9).

Our results suggest several areas for future theoretical and experimental work.

(1) One should calculate all forbidden transition strength in a variety of 
phenomenalogical models.  These calculations should include all incoming p and s-waves 
and all parts of the weak current including vector and axial-charge components.  
The calculations should also include corrections for the spatial dependence of the 
lepton wave functions.

(2) One should calculate meson exchange current contributions for the above transitions.

(3) One should repeat microscopic four-body calculations similar to those
of Carlson et al.\cite{carlson} including all s and p partial waves, all vector and
axial parts of the weak current and the spatial dependence of the lepton wave 
functions. 

(4) It may be useful to compare these calculations to experimental data
for related, nonweak,  reactions.  One should look at radiative capture
${}^3$He(n,$\gamma$)${}^4$He and ${}^3$H(p,$\gamma$)${}^4$He.  Spin-observable data
may provide information on p-wave contributions.  We caution that simple relations
between radiative and weak capture, which have been used in the past, may be
unreliable because of the very different currents involved.   Nevertheless, radiative
capture may still provide useful tests of the models.  In addition, it may be possible
to test calculations of axial-charge strength by comparing to near threshold s-wave
pion production in ${}^3$He(p,$\pi^+$)${}^4$He.  Finally, there could be forbidden 
$0^+\rightarrow 0^+$ Fermi strength.  This might be observable via ${}^3$H(p,$e^+e^-$)
${}^4$He.

(5) Solar neutrino experiments such as SuperKamiokande, SNO\cite{SNO} and 
Icarus\cite{Icarus} should carefully search for hep neutrinos at energies near 14 MeV 
and above.  Although the flux is relatively small there could be a significant number 
of events in an energy region with very low background.  {\it It is important to set
experimental limits on the hep flux that are independent of theory.}

In conclusion, high energy Solar neutrinos can come from the hep reaction 
${}^3$He(p,$\nu e^+$)${}^4$He.  Therefore, the interpretation of measured spectra may 
depend on our understanding of the hep $S$ factor.  The present very small estimate for 
$S$ assumes a pure Gamow-Teller transition that is greatly reduced by cancelations and 
destructive interference from meson exchange currents.   We use a simple model of 
a nucleon moving in optical potentials to show that the axial-charge transition 
${}^3P_0\rightarrow{}^1S_0$ may have comparable strength.  It is important to calculate 
the contribution of other forbidden transitions.  This strength could significantly 
enhance the hep $S$ factor.

Helpful conversations with Hamish Robertson are acknowledged.  This work was supported 
in part by DOE grant:  DE-FG02-87ER40365.

%\eject

%{\bf Figures} %\vskip 5in %\special{bmp:twocolor.bmp x=5in y=4in}

\vbox to 4.5in{\vss\hbox to 8in{\hss 
{\includegraphics{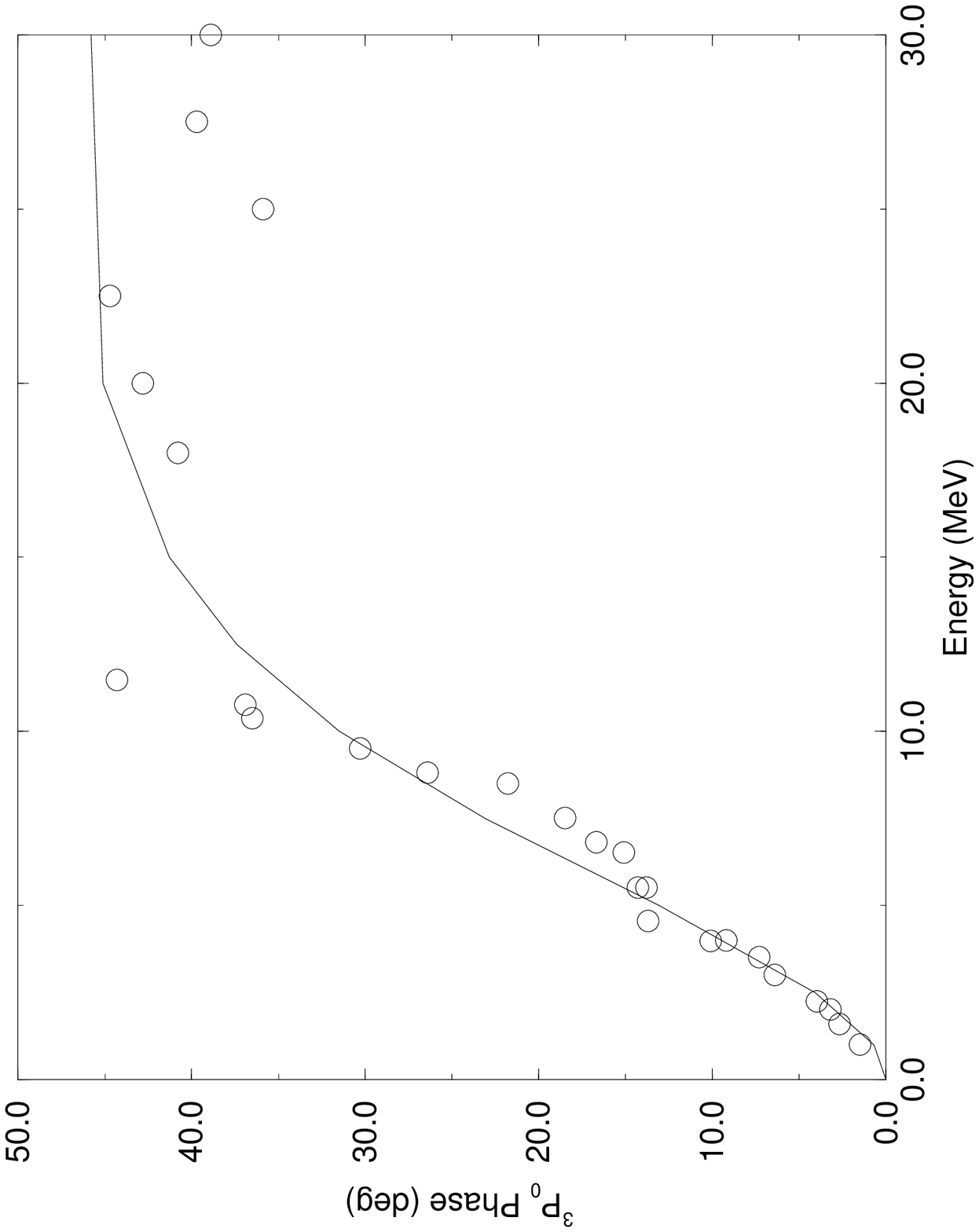}}\hss}} \nobreak {\noindent\narrower{{\bf FIG.~1}.
Phase shift for p+${}^3$He elastic scattering in the ${}^3P_0$ partial wave.  The solid
curve is for the optical potential, Eq. (4), with $V_0=-30$ MeV.  The data is from 
ref.\cite{lowEphases} below 12 MeV and from ref.\cite{highEphases} above 12 MeV.
}} 

\begin{references} 
\bibitem{superK}SuperKamiokande Collaboration, Y. Suzuki, in Neutrino 98, 
Proceedings of the XVIII International Conference on Neutrino Physics and Astrophysics, 
Takayama, Japan, 4-9 June 1998, Y. Suzuki, Y. Totsuka (Eds.).  To be published in 
Nucl. Phys. B (Proc. Suppl.).

\bibitem{bahcall} John N. Bahcall and Plamen I. Krastev, Phys. Lett. B {\bf 436} 
(1998) 243.

\bibitem{escribano} R. Escribano, J.-M. Frere, A. Gevaert and D. Monderen, Phys. Lett.
B {\bf 444} (1998) 397.

\bibitem{BP98} J.N. Bahcall, S. Basu and M. H. Pinsomeault, Phys. Lett. B 
{\bf 433} (1998)1.

\bibitem{carlson} J. Carlson, D. O. Riska, R. Schiavilla and R. B. Wiringa, 
Phys. Rev. {\bf C44}(1991) 619.

\bibitem{schiavilla} R. Schiavilla, R. B. Wiringa, V. R. Pandharipande, J. Carlson, 
Phys. Rev. {\bf C45} (1992) 2628.

\bibitem{werntz1} C. Werntz and J.G. Brennan, Phys. Rev. {\bf 157} (1967) 759.

\bibitem{werntz2} C. Werntz and J.G. Brennan, Phys. Rev. {\bf C8} (1973) 1545.

\bibitem{tegner} P.E. Tegner and Chr. Bargholtz, Astrophys. J. {\bf 272} (1983) 311.

\bibitem{wolfs} F.L.H. Wolfs, S.J. Freedman, J.E. Nelson, M.S. Dewey, G.L. 
Greene, Phys. Rev. Lett. {\bf 63} (1989) 2721.

\bibitem{wervelman} R. Wervelman, K. Abrahams, H. Postma, J.G.L. Booten, and
A.G.M. Van Hees, Nucl. Phys. {\bf A526} (1991) 265.


\bibitem{pions} T.-S.~H. Lee and D.~O. Riska,
Phys.\ Rev.\ Lett.\ {\bf 70}, 2237 (1993).
C.J. Horowitz, H.O. Meyer and D.K. Griegel, Phys. Rev. {\bf C49} (1994) 1337. 

\bibitem{SNO} A. B. McDonald, Proceedings of the 9th Lake Louise Winter Institute, 
A. Astbury et al. (Eds.) Singapore, World Scientific, 1994, p.1.

\bibitem{Icarus} J.P. Revol, in: Frontiers of Neutrino Astrophysics, Y. Suzuki, K. 
Nakamura (Eds.) Tokyo, Universal Academy Press, Inc., 1993, p. 167.

\bibitem{lowEphases} T.A. Tombrello, Phys. Rev. {\bf 138} (1965) B40.

\bibitem{highEphases} J.R. Morales, T.A. Cahill, D.J. Shadoan and H. Willmes, 
Phys. Rev. {\bf C11} (1975) 1905.

\end{references}
\end{document}